\documentclass{emulateapj}
\usepackage{graphicx}    
\usepackage{color}
\usepackage{natbib}

\newcommand{\Msun}{M_{\odot}}
\newcommand{\Mvir}{M_{\rm vir}}
\newcommand{\Rvir}{R_{\rm vir}}

\newcommand{\Mstar}{M_{\rm star}}
\newcommand{\NHI}{N_{\rm HI}}
\newcommand{\Msh}{M_{\rm sh}}
\newcommand{\CF}{f_{c}}
\newcommand{\CFR}{f_{c}(<\hspace{-0.7 ex}R)}

\slugcomment{}

\begin{document}
\title{Observing the End of Cold Flow Accretion using Halo Absorption Systems}
\author{Kyle R. Stewart\altaffilmark{1}, Tobias Kaufmann\altaffilmark{2}, James S. Bullock\altaffilmark{3,4}, Elizabeth J. Barton\altaffilmark{3,4},
  Ariyeh H. Maller\altaffilmark{5},
  J\"urg Diemand\altaffilmark{6},
  James Wadsley\altaffilmark{7}}

\altaffiltext{1}{NASA Postdoctoral Program Fellow, Jet Propulsion Laboratory, Pasadena, CA 91109, USA}
\altaffiltext{2}{Institute for Astronomy, ETH Zurich, CH-8093 Zurich, Switzerland}
\altaffiltext{3}{Center for Cosmology, Department of Physics and Astronomy, The University of California at Irvine, Irvine, CA, 92697, USA}
\altaffiltext{4}{Center for Galaxy Evolution, Department of Physics and Astronomy, The University of California at Irvine, Irvine, CA, 92697, USA}
\altaffiltext{5}{Department of Physics, New York City College of Technology, 300 Jay St., Brooklyn, NY 11201, USA}
\altaffiltext{6}{Institute for Theoretical Physics, University of Zurich, 8057, Zurich, Switzerland}
\altaffiltext{7}{Department of Physics and Astronomy, McMaster University, Main Street West, Hamilton L85 4M1, Canada}

\begin{abstract} {
We use cosmological SPH simulations to study the cool, accreted gas in two Milky Way-size galaxies through cosmic time to $z=0$.
We find that gas from  mergers and cold flow accretion results in significant amounts of cool gas in galaxy halos.  This cool circum-galactic
component drops precipitously once the galaxies cross the critical mass to form stable shocks, $\Mvir = \Msh \sim 10^{12}\Msun$.
Before reaching $\Msh$, the galaxies experience cold mode accretion ($T < 10^5$ K) and show moderately high covering fractions in
accreted gas: $\CF \sim 30-50\%$ for $R<50$ co-moving kpc and $\NHI>10^{16}$ cm$^{-2}$.
These values are considerably lower than observed covering fractions, suggesting that outflowing gas (not included here)
is important in simulating galaxies with realistic gaseous halos.
Within $\sim500$ Myr of crossing the $\Msh$ threshold, each galaxy transitions to hot mode gas accretion,
and $\CF$ drops to $\sim5\%$. The sharp transition in covering fraction is primarily a function of halo mass, not redshift. This
signature should be detectable in absorption system studies that target galaxies of varying host mass, and may provide a direct
observational tracer of the transition from cold flow accretion to hot mode accretion in galaxies.
}
\end{abstract}
\keywords{cosmology: theory --- galaxies: formation --- galaxies: evolution --- galaxies: halos ---  methods: numerical}

\section{Introduction}
\label{Introduction}
Recent advances in galaxy formation theory have emphasized the importance of cool gas accretion onto galaxies:
gas that never shock-heats to the virial temperature of the halo
\citep[e.g.,][]{BirnboimDekel03, DekelBirnboim06, Keres09}.
Under this picture of galaxy formation, the cooling time of gas entering
halos less massive than a critical threshold, $\Mvir=\Msh$, is too short to sustain compressive shocks,
with $\Msh\sim10^{11.5-12}\Msun$.
Galaxies within dark matter halos less massive than $\Msh$
experience ``cold mode'' accretion, since baryonic accretion onto these galaxies is dominated by cool gas.
Galaxies above this transition experience ``hot mode'' accretion, with infalling gas shock-heating to the virial
temperature.
In practice, the precise value of $\Msh$ depends somewhat on definition. In what follows we are interested in the
absolute shut-down mass, when almost no cool accreted gas reaches the central regions of the halo.
Thus, we adopt $\Msh\sim10^{12} \Msun$, motivated by \cite{DekelBirnboim06} for shocks at large fractions of the virial radius
for $Z\sim0.1$ metallicity gas\footnote{\cite{Keres09} find that
less than half of a \emph{galaxy's} gas is accreted from cold mode accretion for
$\Mvir\gtrsim2\times10^{11}\Msun$; this fraction drops to near zero at $10^{12}\Msun$.
Above $\sim10^{12}\Msun$ cold mode accretion is negligible.}.

Unfortunately, there are currently no definitive observational tests to detect cosmological cool gas accretion.
On the contrary, numerous observational studies of cool halo gas around galaxies have emphasized the presence
of gas \emph{outflows}, not inflows \citep[e.g.][]{Steidel96, Shapley03, Weiner09, Steidel10,Rubin11}.
The stark contrast between theory and observations is understandable at high redshift, as gas inflow to
galaxies at $z>2$ is expected to flow along dense filaments, resulting in small global covering fractions \citep{FGKeres10,Kimm10}.
In addition, galaxies at $z\sim2$ are at the peak of cosmological star formation \citep{HopkinsBeacom06}; one might
expect feedback processes to dominate any observational indicator of gas accretion at these epochs.

At lower redshifts star formation rates decline and the gaseous halos of galaxies are observed as quasar absorption systems
\citep[e.g.,][]{BergeronBoisse91,Bowen95,Churchill96,
BartonCooke09,
Chen10,Gauthier10}.
These studies are primarily of metal lines (MgII, CIV, OVI, etc.) though with COS on HST we expect Lyman $\alpha$ observations
to increase.  In this letter, we utilize cosmological hydrodynamic simulations to study cool gas accretion and the possibility
of detecting it as quasar absorption systems.  To correctly produce metal lines in a simulation requires radiative transfer,
metal diffusion and modeling of local ionizing sources; however, we are focusing on the qualitative behavior of halo gas, so we will
instead give results in terms of HI column density calculated in the optically thin limit, without local
sources.  For column densities below the Lyman limit ($2 \times 10^{17}$ cm$^{-2}$) this should be fairly robust, but for higher column
densities we expect a full treatment would lead to quantitative but not qualitative differences.


\section{Simulations and Analysis}
\label{simulation}
\subsection{The Simulations}
Our two simulations utilize separate sets of cosmological initial conditions, with each simulation tracking the evolution of
a roughly Milky Way size dark matter halos until $z=0$.  We refer to these two simulations
as ``Halo $1$'' and ``Halo $2$'', since we primarily investigate the properties of the single most massive galaxy in each simulation.
Halo $1$ has an active merger history until $z\sim1.5$, but subsequently
experiences a relatively quiescent merger history
(a dark matter only simulation of the same initial conditions was performed at very high resolution in
the Via Lactea II simulation of \cite{VL2}
In contrast, Halo $2$ experiences a more quiescent early history,
but has a major merger at $z\sim0.8$. For both simulations, the most massive galaxy has a dark halo mass of
$\Mvir(z=0)=1.4\times 10^{12} \Msun$, and is positioned within a large scale filament of dark matter and gas,
typical for galaxies of this mass.

We use multiple mass particle grid initial conditions generated with the GRAFIC-2 package (\cite{Bertschinger01}) and the best-fit
cosmological parameters of the WMAP three-year data release \citep{WMAP3}:
$\Omega_{M} = 0.238$, $\Omega_{\Lambda} = 0.762$, $H_{0}= 73 $km s$^{-1}$ Mpc$^{-1}$, $n_s=0.951$, and $\sigma_8=0.74$.
We implement the ``zoom in'' technique to properly account for large-scale tidal torques as described in \cite{KatzWhite93},
and each simulation is contained within a periodic box of $40$ co-moving Mpc, with the highest resolution region limited to
a $\sim6$ co-moving Mpc cube.
We use the smoothed particle hydrodynamics (SPH) code GASOLINE \citep{GASOLINE}.
In the highest resolution region of each simulation, the masses of the simulated particles in the initial
conditions are: $(m_{\rm dark}, m_{\rm gas}) = (17,3.7)\times10^5 \Msun$, with a force softening of $332$ pc.

The code assumes a uniform UV background from QSO, implemented following
\cite{HaardtMadau96} and F. Haardt (2002, private communication).
It implements star formation, as well as Compton and radiative cooling,
as described in \cite{Katz96}, calculating the abundance of neutral hydrogen
by assuming  an optically thin ideal gas of primordial
composition and in ionization equilibrium  with the UV-background, treating collisional ionization, photoionization and recombination processes.
We do not include full treatment of radiative transfer.  Because we primarily focus on qualitative attributes of
cool halo gas, and its evolution with time, we do not believe this will significantly impact our results.
We expect that a full treatment would increase the amount of very high density gas, causing a
progressively larger increase to our reported covering fractions, starting at a few times the Lyman limit.

Type II supernovae are modeled using an analytical treatment of blastwaves,
creating turbulent motions in nearby gas particles that
keeps them from cooling and forming stars, as described in \cite{Stinson06}.
This feedback model results in minimal winds of $\sim100$ km/s that mostly affect hot gas
and are more prominent in $\Mvir\lesssim10^{11}\Msun$ halos \citep{Shen10}.
For the mass range considered in this letter, this feedback model does \emph{not}
result in cool outflows that would be detectable through absorption.
The only two free parameters in the star formation and feedback model
have been fixed in order to produce galaxies with a realistic star formation rates,
disk thicknesses, gas turbulence, and Schmidt law over a range in dynamic masses \citep{Governato07}.
We note our feedback model is very similar to those used in recent
simulations that have shown great success in producing realistic disk-type galaxies \citep{Governato09}, as well
as matching the mass-metallicity relation \citep{Brooks07} and the
abundance of Damped Lyman $\alpha$ systems at $z=3$ \citep{Pontzen08}.  We
refer the reader to \cite{Governato09} for a more detailed description of the simulation code.
At each output snapshot of our simulation, we define the virial radius by \cite{BryanNorman98}, noting
that this is a fairly typical definition used in $N$-body simulations of dark matter substructure \citep[e.g.][]{Stewart08}.

While a detailed morphological analysis of our simulated galaxies is beyond the scope of this letter,
visual inspection shows that both galaxies are disks before they reach the transition mass for hot mode accretion.
After this transition, Halo $1$ remains roughly disk-like but grows a massive bulge as hot gas cools from the
halo, similar to the most massive galaxy in \cite{Governato07} and \cite{Brooks09}.
Due to the major merger in Halo $2$ just prior to reaching the transition mass, it develops into a more spheroidal
galaxy.  We note that previous works that have produced disk galaxies at $z=0$, even for galaxies with major mergers at $z<1$,
typically involve halos less massive than those we study here (e.g., $7\times10^{11}\Msun$ in \cite{Governato09}), which
may be the cause of these morphological differences.

\subsection{Analysis}
To study the cool halo gas in our simulated galaxies, we analyze $794$ regularly spaced sight lines
within $100$ co-moving kpc, for three orthogonal orientations of each galaxy.
For each sight line, we compute the column density of neutral hydrogen per unit velocity, as a
function of velocity along the line of sight, mimicking the practice of searching for absorption gas along a quasar line of sight.
We do this by dividing each sight line into spatial bins of order the size of the spatial
resolution of the simulation.  For each spatial bin, each gas particle with a smoothing length that intersects the line of sight
contributes to the total mass density of the bin.  For each of these particles, the contribution of the spline kernel
to the column density of each bin that overlaps the kernel is integrated along the line of sight.  The velocity of each bin is
the mass-weighted velocity of particles contributing to the total column density of neutral hydrogen.

\begin{figure*}[tbh!]
  \includegraphics[width=0.49\textwidth]{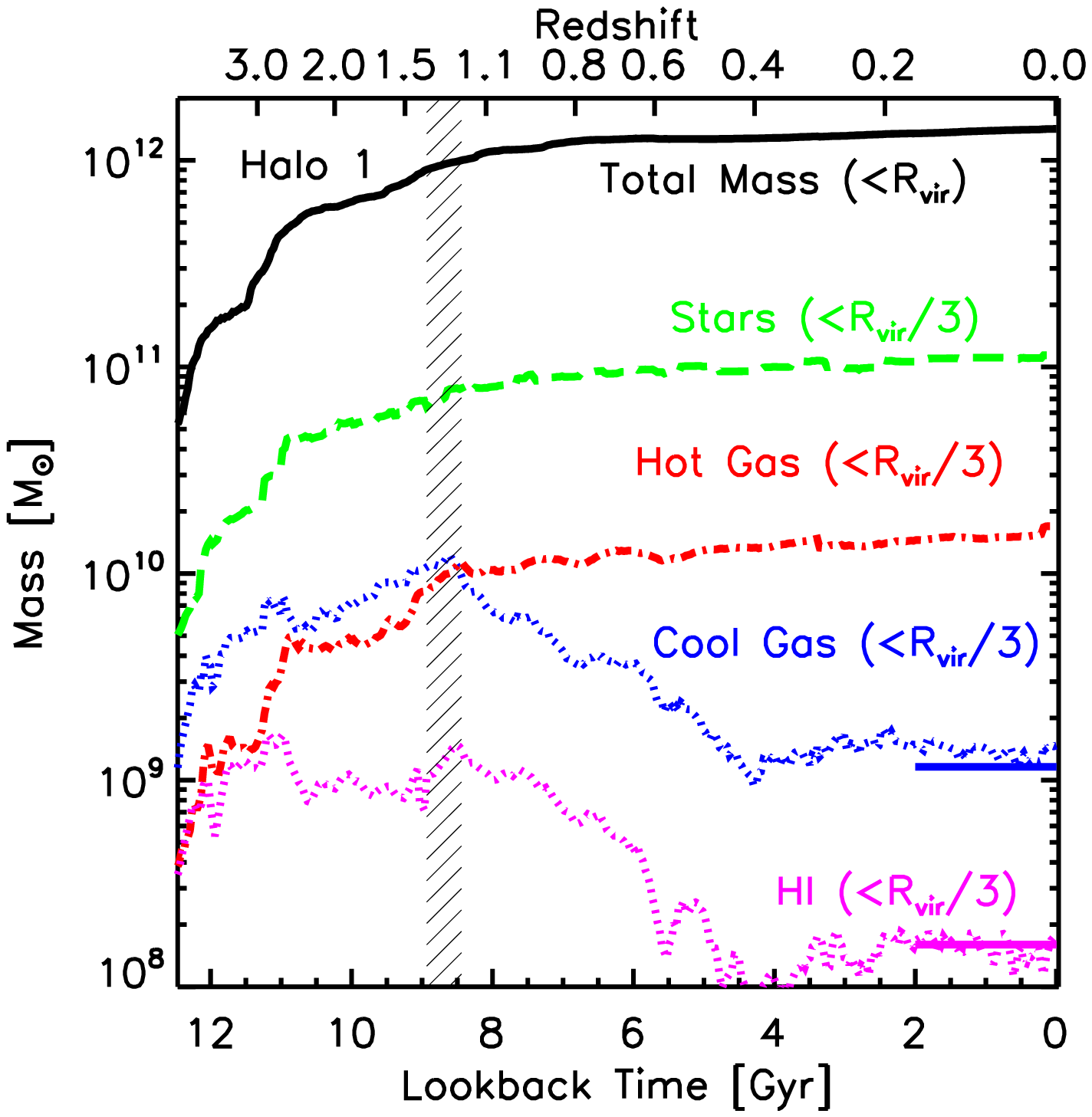}
  \includegraphics[width=0.49\textwidth]{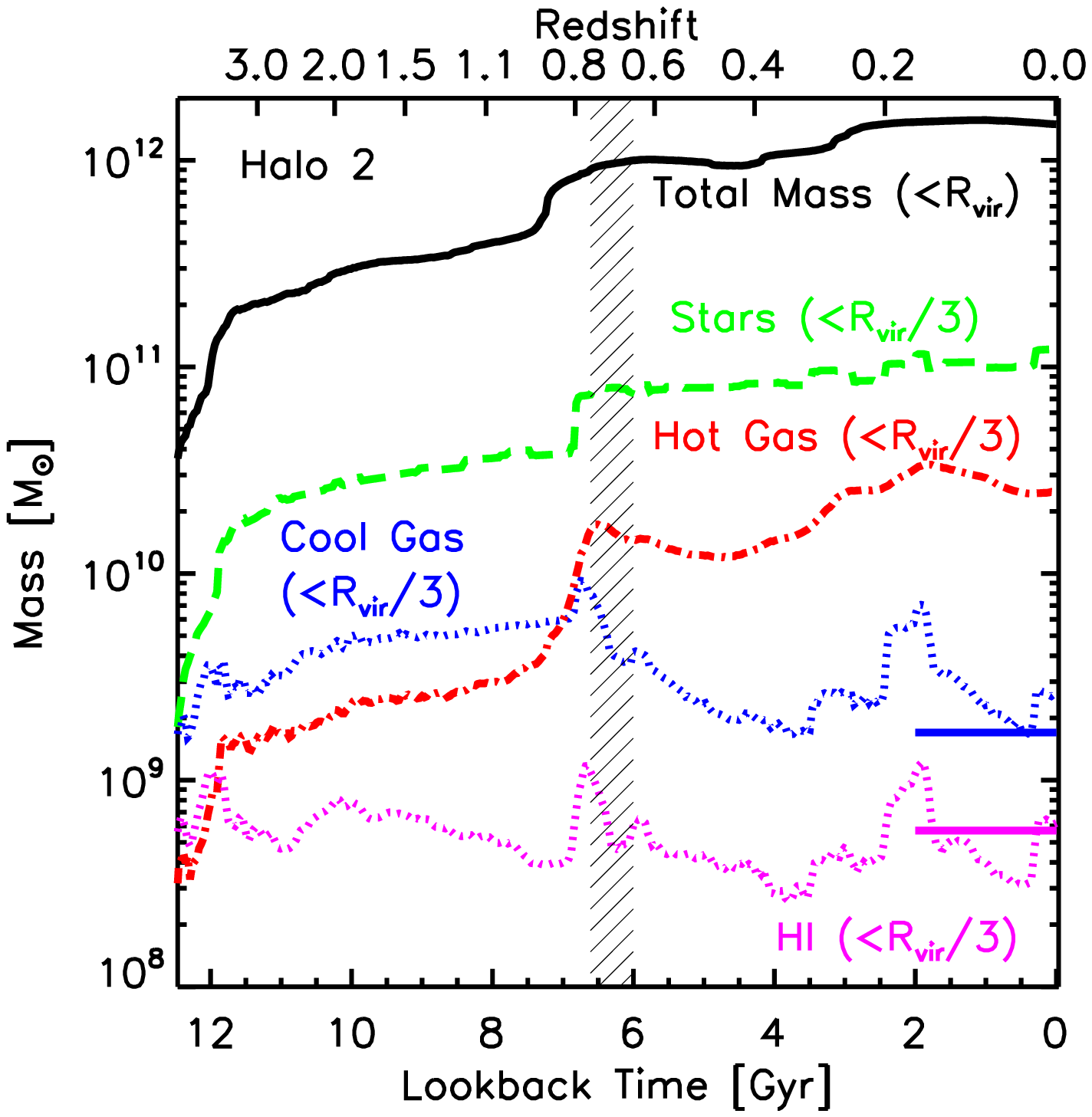}
 \includegraphics[width=0.49\textwidth]{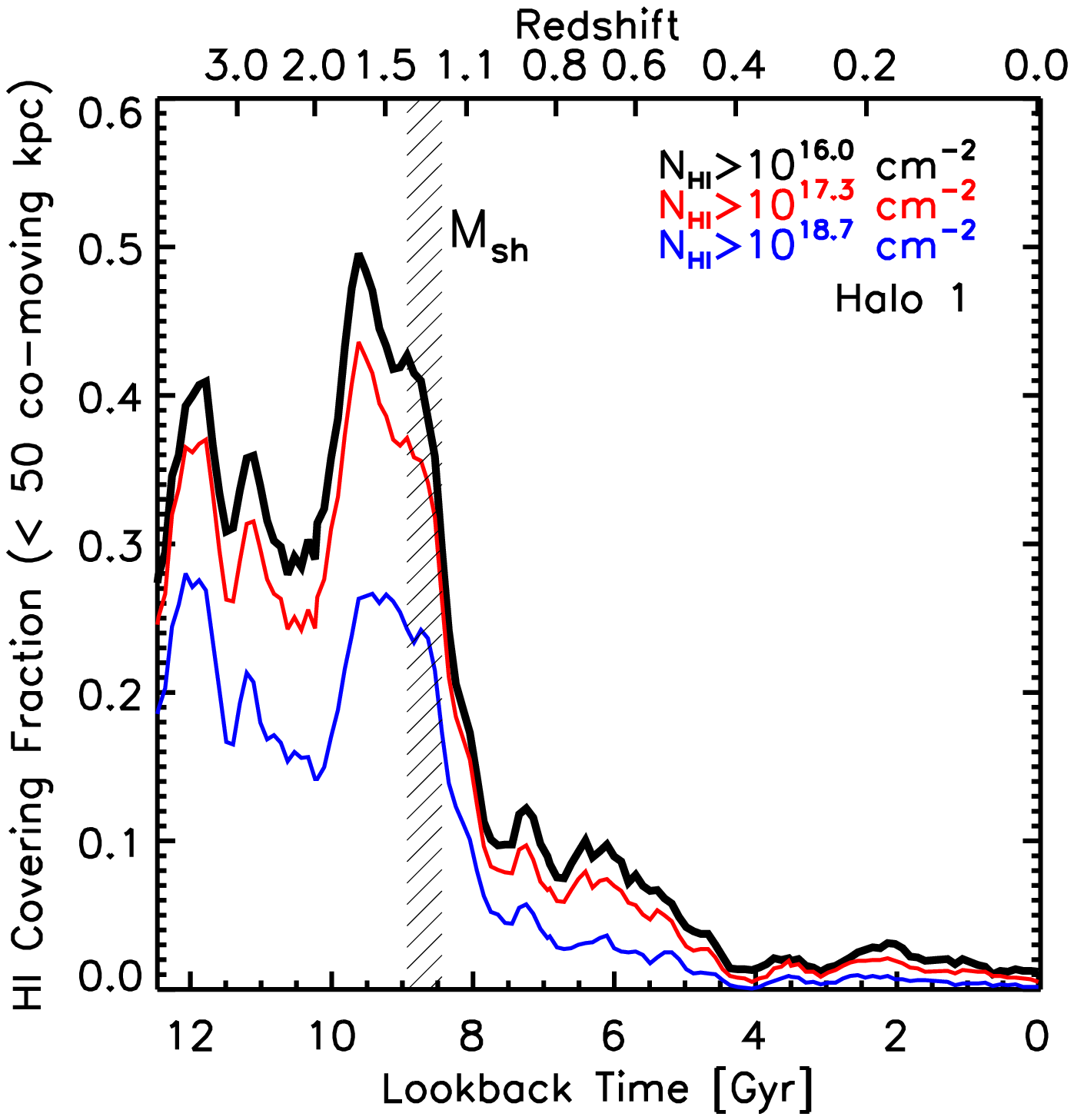}
 \hspace{1em}
 \includegraphics[width=0.49\textwidth]{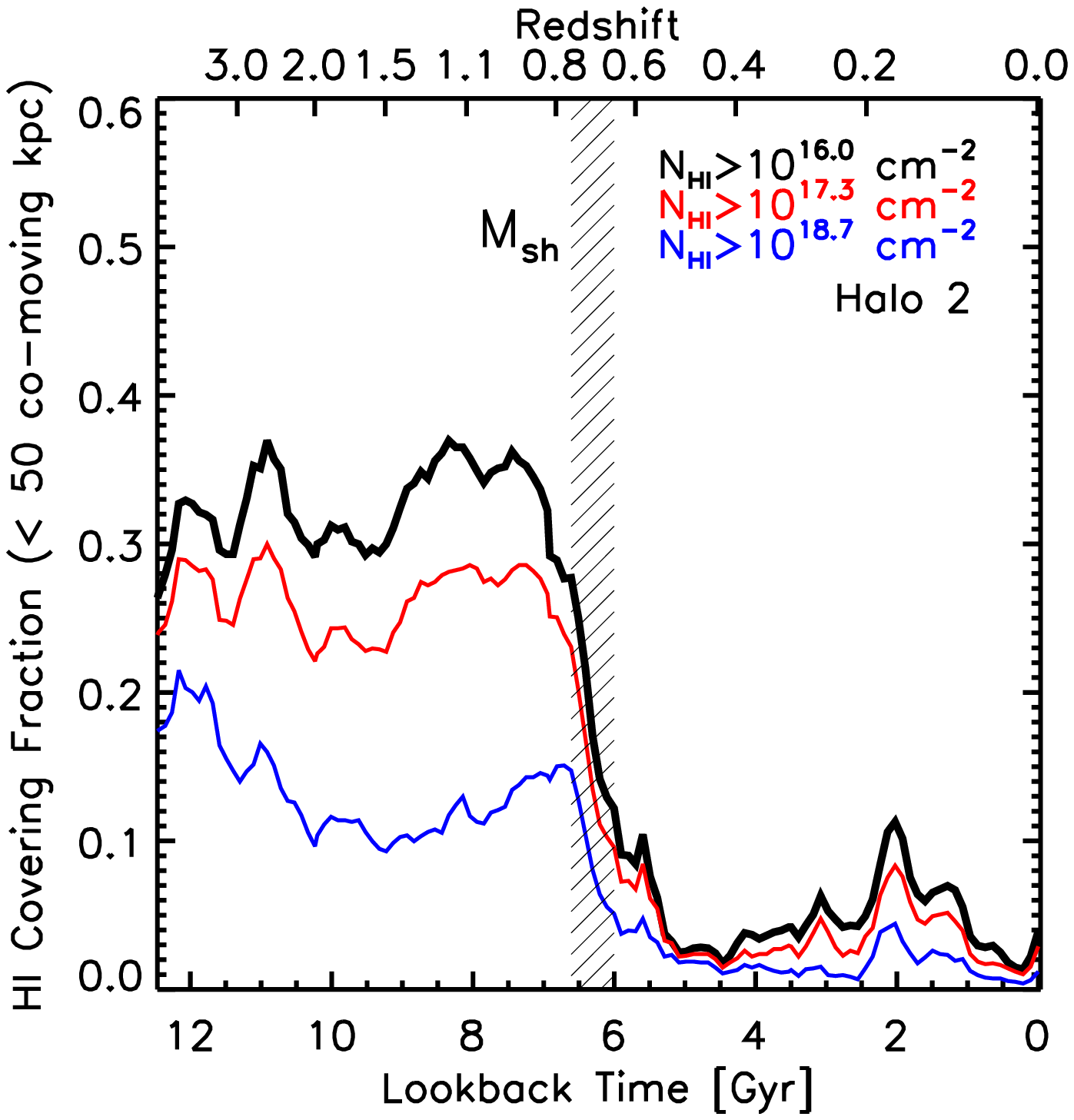}
 \caption{Masses and covering fractions of each galaxy versus look-back time.  The left panels show Halo $1$, while the right panels show Halo $2$.
 \emph{Top:} the total virial mass (solid black), and baryonic masses within $R<\Rvir/3$ ($\simeq 100$ kpc at $z=0$ for halos of this mass).
  Stellar mass, hot gas mass ($T>10^5$ K), cool gas mass ($T\leq10^5$ K) and neutral hydrogen are given by the dashed green, dash-dot red, dotted blue,
  and dotted magenta lines.  The thick horizonal lines on the right side of each plot show the cool gas mass and neutral hydrogen contained within
  the central galaxy and massive satellite galaxies at $z=0$.
  \emph{Bottom:} covering fraction of neutral hydrogen within $R<50$ co-moving kpc as a function of time.  Black, red and blue lines represent
 minimum column densities $\NHI>10^{16.0,17.3,18.7}$ cm$^{-2}$, respectively.
 The vertical hashed bar in each panel shows when the galaxy transitions from cold mode to hot mode accretion (see \S\ref{coldhotmode}).
}
\label{vstime}
\end{figure*}

We define the \emph{covering fraction} of accreted neutral hydrogen, $\CFR$, as the total fraction of
sight lines (within a radius $R$ from the center of the galaxy) for which $\NHI>10^{16.0-18.7}$ cm$^{-2}$.
For $>95\%$ of cases, the full width half maximum spread in velocities of absorbing
sightlines\footnote{Because our feedback model does not produce cool gas outflows, our absorption sightlines
are never offset from the systemic velocity by more than $1,000$ km/s.}
is $<110$ km/s.
We choose these column density limits in order to probe different types of cool
gaseous structures that may exist in the halos of galaxies, noting that they correspond to a plausible minimum
threshold for halo gas detection via Mg$\,$II absorption \citep{Rigby02, Churchill00}.

Some past studies of metal line absorption systems associated with galaxies
suggest that there may be a critical radius, outside of which cool gas clouds cannot form
\citep[e.g.][]{Chen01, Chen10}, while a study by \cite{Steidel10} suggests that $\CFR$ decays as a power law in $R$.
Based on the typical extent of neutral hydrogen in our galaxies,
we present covering fractions in terms of three fixed radii: $R<30, 50, 100$ co-moving kpc
\cite[for a detailed exploration of the radial dependence of $\CFR$, see][]{Stewart11b}.
We include the galaxy disk in our analysis, but its contribution to the covering fraction is relatively minor.
For $R<30,50, 100$ co-moving kpc, $\CF$ for the disk is $<10,4,1\%$, respectively.

\section{Galaxy Growth and Evolution}
\label{galaxygrowth}

Figure \ref{vstime} shows the mass accretion history of our galaxies, as well as the covering fraction
of each galaxy over time (Halo $1$ on the left, Halo $2$ on the right).
In the top panels, the virial mass is given by the solid black line, while other curves give baryonic masses within
$\Rvir/3$ ($100$ kpc at $z=0$).  The stellar mass, hot gas mass ($T>10^5$ K),
cool gas mass ($T\leq10^5$ K), and neutral hydrogen mass are given by the dashed green, dot-dashed red, dotted blue, and dotted
magenta lines, respectively.
Halo $1$ experiences several large mergers at $z>1.5$, with a relatively quiescent growth after $z=1$, while Halo $2$
has a more quiescent early growth ($z<1$), but experiences a major merger at $z\sim0.8$, and another moderately large merger
at $z\sim0.2$. These signatures, especially the late-time mergers for Halo $2$, can be seen by the sharp
increases in various curves in the top panels.
Note that most of the halo gas within $R<\Rvir/3$ is cool until the halo grows to $\Mvir\sim10^{12}\Msun$.
(See \S\ref{coldhotmode} for more on this transition.  The vertical hashed bar in each panel shows the range
where $\Mvir=0.9-1.0\times10^{12}\Msun$ for each galaxy.)
For proper context, we also include the cool gas mass enclosed within the main galaxy ($R<10$ kpc)
or in massive satellite galaxies at $z=0$, represented by the thick horizontal lines at the right side of each plot.
It is clear, then, that the vast majority of cool gas is associated with galaxies at late times and is not
spread throughout the halo.  In contrast, when the galaxy first reaches the transition mass, the cool
gas in the galactic disk is only $\sim10^{9}\Msun$, and does not dominate the cool gas mass within $\Rvir/3$.

\begin{figure*}[tbh!]
  \includegraphics[width=0.49\textwidth]{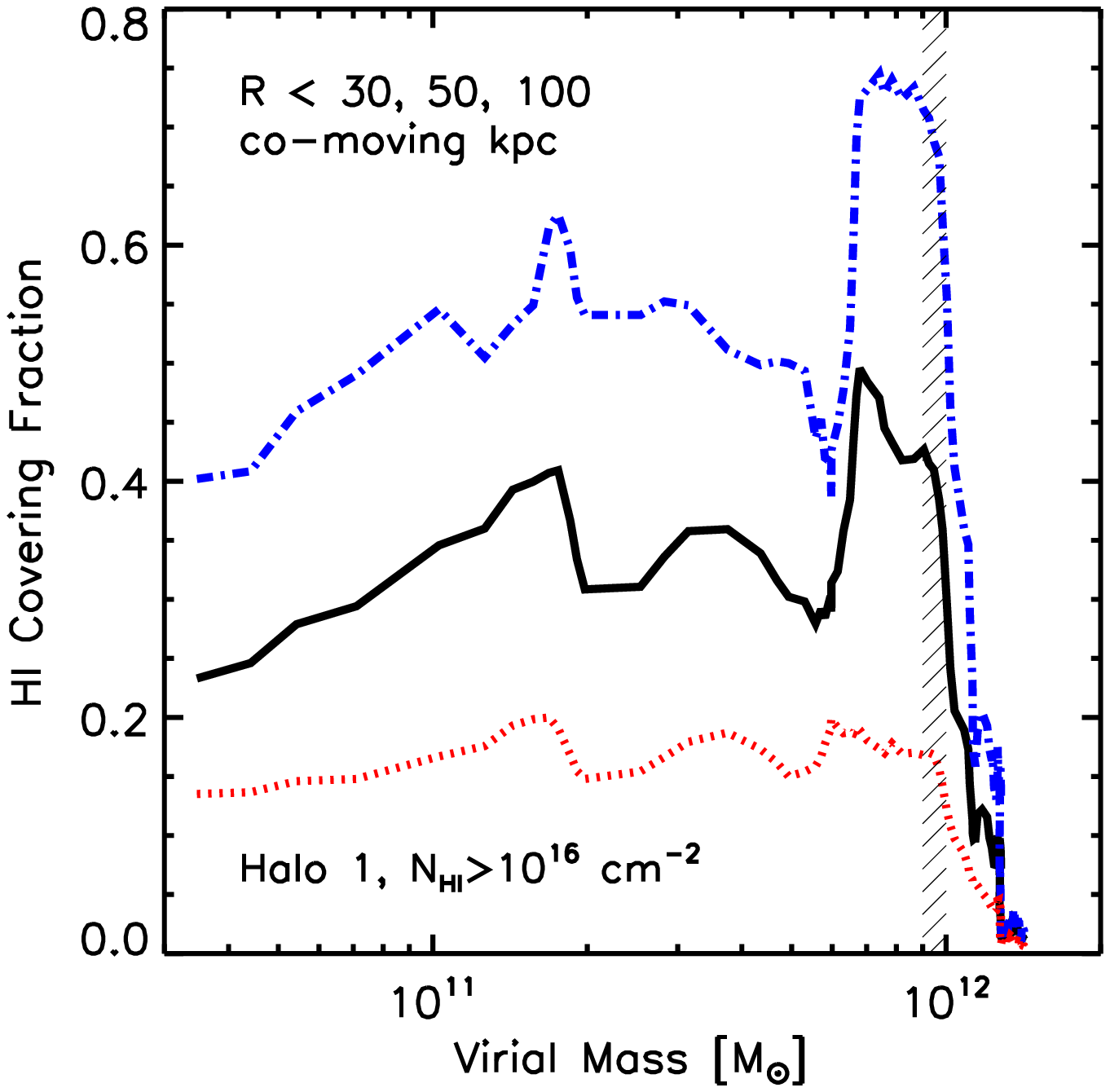}
  \includegraphics[width=0.49\textwidth]{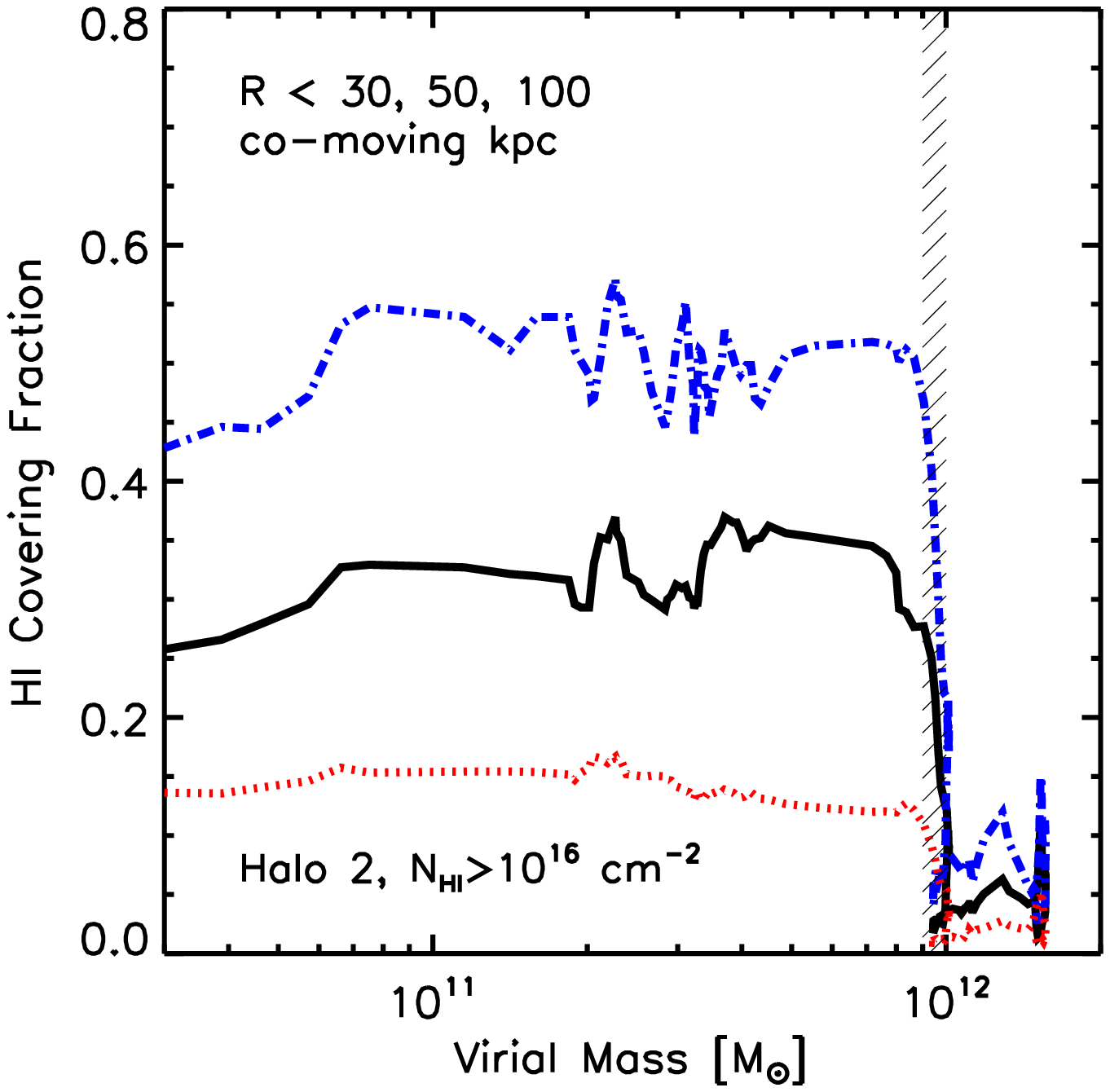}
 \caption{Covering fraction of neutral hydrogen as a function of the halo virial mass.
 The left and right panels show Halo $1$ and Halo $2$.  Different lines
 correspond to outer radii of $R<30, 50, 100$ co-moving kpc (dot-dashed blue, solid black, dotted red, respectively).
 Note the drastic transition in covering fraction at the cold/hot mode transition
 of $\Mvir=\Msh\sim10^{12}\Msun$, which occurs at $z\sim1.3$ for Halo $1$ and $z\sim0.8$ for Halo $2$.
 }
\label{vsmass}
\end{figure*}

The bottom panels of Figure \ref{vstime} show the time evolution of $\CF$($R<50$ co-moving kpc) for
each galaxy, with different curves representing different column
densities\footnote{Note that the higher column density lines
will not be correct, in detail, because we do not have full treatment of radiative transfer in our analysis.  Our
estimates are likely too low for higher density gas.}:
$N_{HI}>10^{16.0}$ cm$^{-2}$ (thick black), $N_{HI}>10^{17.3}$ cm$^{-2}$ (thin red), and $N_{HI}>10^{18.7}$ cm$^{-2}$ (thin blue)
In general, both galaxies show remarkably flat evolution of covering fraction with time (for $\Mvir<\Msh$),
with $\CF\sim30-40\%$ for $z<4$.
In detail, each galaxy's covering fraction varies with recent merger history, as demonstrated
by the sudden spikes and dips in Figure \ref{vstime}.  These variations are primarily caused by mergers, which
play an important role in cumulative gas accretion onto galaxies as well as halos
\citep[especially at $z>1$, when mergers are more frequent and more gas-rich than at late times; see e.g.,][]{Stewart09a,Stewart09b}.
We note that \cite{Kacprzak10} have carried out a similar investigation to what we present here,
and also found an accretion-related origin for many Mg$\,$II absorption
systems. In addition, \cite{Kacprzak07} found a correlation between galaxy asymmetry and the presence of halo gas,
again suggesting a link between gas accretion and halo absorption systems.
Reassuringly, our covering fractions at high redshift are in qualitative agreement with
a high-redshift simulation by \cite{FGKeres10}, utilizing a different SPH code and
including treatment of radiative transfer.  For similar mass galaxies as our own,
they quote an average covering fraction of $\sim20-30\%$ at $z=2-4$ ($R<\Rvir/2$ and
$\NHI>10^{17.2}$ cm$^{-2}$).

We emphasize that these moderate covering fractions are a result of gas accretion only.  Our feedback model does
not produce significant outflows.  We do this both to achieve a robust result
(the physical mechanism behind galaxy outflows is largely uncertain) and to focus on signatures of
cold flow accretion.  Since galactic outflows should populate the halo
with additional cool gas, our results serve as a lower limit on the true covering fractions of galaxies.

\section{Cold Flows and Covering Fractions}
\label{coldhotmode}
Observed covering fractions typically vary from $\sim50\%$ \citep[e.g.,][]{TrippBowen05, ChenMulchaey09}
to near unity \citep[e.g.,][]{Chen01,ChenTinker08}, but there are indications that more massive galaxies (LRGs)
may have systematically lower values \citep[$\sim10-15\%$,][]{Gauthier10, BowenChelouche10}.  Despite this variation in the
literature, our findings are significantly lower than those observed, suggesting that outflows
are a required component in producing galaxies with realistic gaseous halos.
Still, the contribution from accreted gas presented here is significant, so that the sudden drop in $\CF$ (Figure \ref{vstime})
should be observable in real galaxies.

Keeping in mind the theoretical framework of cold and hot mode accretion (see \S\ref{Introduction}),
Figure \ref{vsmass} shows the covering fraction as a function of halo virial mass rather than time, for
three choices of outer radii: $30, 50,$ and $100$ co-moving kpc (dot-dashed blue, solid black,
dotted red lines, respectively).  Regardless of the radius used,
$\CFR$ declines sharply at $\Mvir=\Msh$, when cold mode accretion ends for each
galaxy\footnote{To emphasize that a precise value of $\Msh$ is uncertain,
the vertical hashed bar in each panel shows a small range of values, $\Msh=0.9-1\times10^{12}\Msun$.
}.

A preliminary analysis suggests the timescale for cool halo gas to fall onto the central galaxy is short
(Stewart11c et al.~in preparation); thus the reason for this sharp decline in covering fraction becomes clear.
Once the galaxy crosses the threshold into hot mode accretion,
the existing cool halo gas quickly sinks onto the galaxy and cannot be replaced by subsequent gas accretion.
A galaxy in this hot mode accretion can only maintain a cool gaseous halo by mechanisms other than direct accretion.
For example, outflowing winds may be capable of re-distributing cool gas from the galaxy into the halo.
Alternatively, galaxy halos can form instability-triggered clouds,
resulting in cool gas that is neither freshly accreted nor outflowing
\citep[e.g.][]{MallerBullock04, Kaufmann09, KeresHernquist09}\footnote{While
our resolution does not allow us to directly simulate these clouds, we do not expect high covering fractions
due to cloud fragmentation, for standard hot gas profiles \citep{Kaufmann09}.}.

This picture of gaseous halo formation has direct consequences on observations of halo absorption systems.
We predict that galaxies above the transition mass have drastically smaller covering fractions within
a fixed co-moving radius, with respect to \emph{accreted} cool gas.
While we have presented a small range in values for $\Msh$, different models vary in detail.
Typically, $\Msh\sim 10^{11.5-12.5}\Msun$, corresponding to central galaxies with stellar masses
$\Mstar\sim10^{10.2-10.8}\Msun$ at $z=0$, or $\Mstar\sim10^{9.8-10.8}\Msun$ at $z=1$
\citep[based on abundance matching, e.g.,][]{cw09}.
If moderately high covering fractions are found to exist for more massive galaxies (which are not
undergoing gas-rich mergers), that gas is likely the result of outflows, and should show
distinct differences in metallicity and kinematics when compared to freshly accreted cool gas
\citep[for an observable kinematic signature of cool accreted gas, see][]{Stewart11b}.
We note that \cite{TinkerChen08} found observational evidence from
correlation functions that supports this model of a cold/hot mode transition influencing
the presence of halo absorption systems.

\section{Conclusion}
\label{conclusion}
We have used two high-resolution, cosmological SPH simulations as a tool for studying the cool gaseous halos of galaxies.
By creating mock observation sight lines, representing cool gas detection via
absorption, we present the covering fraction of neutral hydrogen in our galaxies over time.
Our primary results are summarized as follows.

\begin{enumerate}

\item To first order, the covering fraction of \emph{accreted} cool gas is relatively stable at $\CF\sim 30-40\%$
    ($R<50$ co-moving kpc, $\NHI>10^{16}$ cm$^{-2}$), as long as the galaxy continues to accrete cool gas from the cosmic web.

\item As soon as our simulated galaxies cross the threshold between cold mode accretion and hot mode accretion
    (when the halo is massive enough to support stable shocks at a large fraction of the virial radius, $\Mvir\sim10^{12}\Msun$)
    the lack of cool accreted gas results in a suppression of cool halo gas.
    Within $\sim500$ Myr of reaching this threshold mass, the covering fraction drops from $30-50\%$ to $5-10\%$.
    A transition this sharp should be directly observable via metal line absorption system studies.

\end{enumerate}

We have used a feedback model without cool gas outflows here, focusing on galaxy halo properties that are a natural
consequence of cosmological gas accretion in LCDM.  However, observations have shown that galaxy outflows are an
abundant phenomenon, and likely play an important role in shaping the properties of cool gaseous halos around
galaxies.
We believe future work in comparing observations to a variety of
simulations with different feedback models would prove a valuable tool in testing theoretical models of galaxy formation,
as well as understanding the underlying nature of galaxy halo observations.

\acknowledgements
The simulations presented here were run on the Cosmos cluster at JPL, and the Greenplanet cluster at UCI.
KRS thanks all those who
commented on a preliminary draft of this letter.
This research was carried out at the Jet Propulsion Laboratory, California Institute of Technology,
under a contract with the National Aeronautics and Space Administration.
KRS is supported by an appointment to the NASA Postdoctoral Program at the Jet Propulsion Laboratory,
administered by Oak Ridge Associated Universities through a contract with NASA.
JSB and KRS were partially supported by NASA grant NNX09AG01G.
TK and JD were supported by the Swiss National Science Foundation (SNF).
Copyright 2010. All rights reserved.

\bibliography{corot}

\end{document}